\documentclass[11pt,twoside]{article}   
\usepackage{latexsym} 
\usepackage{mathptm}
\usepackage{amsmath}
\usepackage{amssymb}
\usepackage{graphicx}

\def\beqn{\begin{eqnarray}}
\def\eeqn{\end{eqnarray}}

\begin{document}
 
\title{Comment on the black hole firewall}
\author{Sabine Hossenfelder \thanks{hossi@nordita.org}\\
{\footnotesize{\sl NORDITA, Roslagstullsbacken 23, 106 91 Stockholm, Sweden}}}

\date{}
\maketitle
\vspace*{-1cm}
\begin{abstract}
Recently, it has been argued that black hole complementarity is inconsistent by
showing that, for an infalling observer, it would lead to the existence of a 
firewall near the black hole
horizon, thereby violating the equivalence principle. If true, this would necessitate
to give up on at least one of the postulates of black hole complementarity. In
this comment I want to address an additional assumption that went into the
conclusion, that the early outgoing Hawking radiation is entangled with the
late radiation. 
\end{abstract}


In a recent paper \cite{Almheiri:2012rt} Almheiri, Marolf, Polchinski and Sully (hereafter AMPS) argued
that black hole complementarity \cite{Susskind:1993if} is intrinsically inconsistent. This argument has attracted some
attention because black hole complementarity is presently one of the most popular, maybe the most popular,
solution to the black hole information loss problem. An inconsistency in the assumptions it is based
on would require us to reconsider what we believed to have understood. So far, the majority of follow-up works
argued against the existence of firewalls 
\cite{Nomura:2012sw,Mathur:2012jk,Chowdhury:2012vd,Bena:2012zi,Banks:2012nn,Ori:2012jx}.
The case for firewalls has been made in \cite{Susskind:2012rm}, and in \cite{Bousso:2012as} it has been 
argued that complementarity is not enough. 

The purpose of this brief comment is
to point out an additional assumption that went into the AMPS argument and speculate on the consequences
of relaxing it. 

Before we begin, let us summarize the postulates that are claimed to be inconsistent:
 
\begin{itemize}
\item[P1] Unitarity of black hole evaporation: There exists a unitary $S$-matrix which describes
the evolution from infalling matter at $I^-$ to outgoing Hawking-radiation at $I^+$.
\item[P2] Validity of semi-classical approximation: Outside the stretched horizon of a black hole with 
mass $M$ much above the Planck mass, and outside the Planckian curvature regime, physics
can be described to good approximation by a set of semi-classical field equations.
\item[P3] Statistical interpretation of black hole entropy: To a distant observer, the black hole
appears to be a quantum system with discrete energy levels and entropy $S(M)$, where $S(M)$ is the Bekenstein-Hawking entropy.
\item[P4] The equivalence principle: The effects of gravitational fields are locally indistinguishable from acceleration in flat space.
\end{itemize}

I've taken the liberty to slightly modify these postulates. Note in particular that P4 is much stronger
than the original version, which only states that a freely falling observer who crosses the horizon
does not notice anything `out of the ordinary'. P4 has also been dubbed the `no drama' assumption.
I also added that one expects the semi-classical approximation to break down at high
curvature.

Since an infalling observer should not notice anything out of the ordinary, information must be
able to fall into the black hole with him. Yet, according to P1, the information must also 
be able to make it to an observer at $I^+$. That both are possible without either observer being able to measure
conflicts with P4, which in particular forbids them witnessing a cloning of information, requires that the information 
falling with the observer into the black hole is
`complementary' to the one outside, but no contradiction can occur because there is no simultaneous
description for information inside and outside the black hole. It was shown in \cite{Susskind:1993mu}
that experiments that would allow observers to compare measurements and pin down contradictions 
cannot be carried out without taking into account (unknown) Planck scale effects. Thus, complementarity
seems to work. It also fits nicely with the AdS/CFT correspondence, which probably explains its
popularity.

{\sc AMPS} deliver two arguments. The first argument, in brief, works as follows. Consider an 
observer at $I^+$ who measures the
Hawking radiation. According to P1, the state he measures must be pure. If the early radiation is entangled
with the late radiation, then the more the observer measures from the early radiation, the better
he knows what has to come later. In particular, he knows that tracing out the early radiation will
produce a mixed state for the late radiation. And, if enough time has passed, the observer can 
use the early radiation to construct
a projection operator on a single mode. In this step P3 enters, because without the statistical
interpretation of the entropy, information could be retained until the Planckian phase. 

This single mode can be traced back from the observer at $I^+$ to the near horizon
region, thereby carefully avoiding to come close to the stretched horizon. According to P2
nothing unusual can happen while tracing back this mode. But the single mode
is very different from the state one usually has, in the normal Hawking-case. Normally,
the state of the late radiation at $I^+$ is a thermal mixture. When traced back to the near 
horizon region, this normal state yields vacuum for the infalling observer, up to modes with 
wavelengths of the order of the Schwarzschild radius (which are not forbidden by P4), and the
infalling observer does not notice anything unusual. Tracing back the
single mode on the contrary does not give anything close by the vacuum state of the
infalling observer. Instead, due to the blueshift from the tracing back, he measures 
excitations of high energy, though still below the Planck energy. More pictorially, he burns (though
this might not be visible to anyone who does not actually cross the horizon). This then 
is in conflict with P4, which means the postulates are not consistent.

The second argument {\sc AMPS} deliver is that entropy subadditivity is violated
for three regions they consider: $A$ -- the early outgoing Hawking modes, $B$  -- the late 
outgoing Hawking mode from above, and $C$ -- the interior partner mode of $B$. If the
black hole is old enough, $B$ must have begun to decrease the entropy of the
radiation at $I^+$, so $S_{AB} < S_{A}$. P4 is argued to imply that the interior
mode is entangled with the exterior mode, so $S_{BC} = 0$ and $S_{ABC} = S_A$.
Taken together, the requirement of subadditivity
\beqn
S_{AB} + S_{BC} \geq S_B + S_{ABC} 
\eeqn
can be simplified to $S_A \geq S_B + S_A$ which is violated since $B$ is entangled
with $A$ and $S_B > 0$ after tracing out $A$. 

Let us then note that both of these arguments relied on an assumption which
does not a priori follow from P1 - P4, that the early radiation is
entangled with the late radiation. If it is not, then one cannot construct 
a projection operator on a single highly energetic mode in the late radiation,
and not trace it back to the horizon. There's thus no firewall. And if $A$ and
$B$ are separately pure, there's no issue with the entropy subadditivity either.

That the early radiation is entangled with the late radiation seems
a natural assumption to make. It basically says that all the partners of the
early Hawking particles, those that would normally fall into the singularity
and get lost, should come out as late as possible because that is when
new quantum gravitational effects most plausibly occur. However, this
doesn't follow in any obvious way from the postulates. It seems possible,
for example, that the partners of the early radiation come out already
by the time $E \sim L$ (where $E$ and $L$ are the dimension of the
Hilbert spaces of the early and late radiation) and the late radiation is 
then a sequence of pure states. 

In fact, one could turn the argument around
and say that if one wants to avoid the firewall, then the early radiation
should not be entangled with the late radiation.

Let us note another consequence of the {\sc AMPS} argument then. When
the late radiation is not the normal Hawking mixture (P1) and tracing it
back to the horizon should not result in
highly energetic modes (P4), then the state that the infalling observer finds
in the near horizon region can still not be identical to the normal vacuum
state. If it was, then the observer at $I^+$ would not obtain any information.

This means
there must be various states that, in the near horizon region,
are locally indistinguishable and are all without highly energetic excitations. 
These states can only begin to differ on length scales larger than the 
Schwarzschild radius. One of them gives the normal thermal Hawking state
at $I^+$. The other ones carry information that was presumably retained
by the stretched horizon. An accelerating observer at constant
radius would similarly locally see thermal Unruh radiation and a 
mixed state, regardless of what the state at $I^+$. 

This shouldn't come as much of a surprise because P4 means
basically that there is `nothing' at the horizon, while P1 means that this
`nothing' still has to be able to carry information. The vacuum must
thus be locally degenerate and be able to encode information without
energy being physically present. Note that this is just a consequence
of taking the postulates and combining them with the {\sc AMPS} argument
for tracing back the modes. 

To be very clear: This is not to say that there is no firewall. It is
just to say that if the early radiation is not entangled with
the late radiation, the {\sc AMPS} argument does not work. This doesn't mean
the conclusion is wrong, just that the case needs to be considered
separately. Maybe it also turns out to be inconsistent. Neither is
this to say that black hole complementarity is right. For even if
the postulates are consistent, it remains unclear just exactly how 
the information is supposed to get into the outgoing radiation. 

One of the requirements for Hawking's argument that has
been repeatedly questioned \cite{Lowe:2006xm,Giddings:2007pj} is the 
use of local effective field theory in the near horizon area. If it breaks down,
and non-local effects can make themselves noticeable, this could
offer a mechanism for information to escape. Unfortunately,
the exact reason why it should break down has so far remained
elusive. 

If local effective field theory is to break down, it
is considerably more plausible that this would happen in the
strong curvature regime, thereby invalidating the argument that 
stable or quasi-stable black hole remnants cause a pair-production
problem which relies on effective field theory \cite{Hossenfelder:2009xq}. 
If one is to give up effective field theory anyway, then this seems to be 
the less radical way to do it. Note however that storing information
until the black hole has shrunk to Planck size requires to give
up the statistical interpretation of the Bekenstein-Hawking entropy.

\section*{Acknowledgements}

I thank Steve Giddings, Don Marolf, Joe Polchinski, Stefan Scherer, Lee Smolin and L\'arus Thorlacius
for helpful discussion.

\end{document}